\documentclass[useAMS,usenatbib]{mn2e}
\usepackage{latexsym,graphicx,natbib}

\newcommand\simless{\mathbin{\lower 3pt\hbox
   {$\rlap{\raise 5pt\hbox{$\char'074$}}\mathchar"7218$}}} 
\newcommand\simgreat{\mathbin{\lower 3pt\hbox
   {$\rlap{\raise 5pt\hbox{$\char'076$}}\mathchar"7218$}}} 

\newcommand\kms{{\rm\,km\,s^{-1}}} 
\newcommand\kpc{{\rm\,kpc}} 
\newcommand\masyr{\rm\,mas\,{yr^{-1}}} 
\newcommand\msun{\rm\,M_\odot}
\newcommand\rsun{\rm\,R_\odot}
\newcommand\msmbh{M_{\rm BH}}
\newcommand\mimbh{m_{\rm BH}}

\title[The ejection of the hypervelocity star]
      {Three-body encounters in the Galactic centre: 
	the origin of the hypervelocity star SDSS J090745.0+024507}

\author[Gualandris, Portegies Zwart, Sipior]
       {
	 Alessia Gualandris, $^{1}$\thanks{E-mail: alessiag@science.uva.nl} 
    	 Simon Portegies Zwart$^{1}$
	 and Michael S. Sipior$^{1}$\\
	 $^1$    Astronomical Institute 'Anton Pannekoek',
	 and Section Computational Science,
	 University of Amsterdam,
	 the Netherlands \\
       }

\begin{document}

\date{Accepted 2005 July 14.  Received 2005 July 6; in original form 2005 May 4}
\pubyear{2005}

\maketitle

\begin{abstract}
Hills (1988) predicted that  runaway stars could be accelerated to 
velocities larger than $1000\kms$ by dynamical encounters with 
the supermassive black hole (SMBH) in the Galactic center.
The recently discovered hypervelocity star SDSS J090745.0+024507 
(hereafter HVS) is escaping the Galaxy at high speed and could be the first
object in this class. 
With the measured radial velocity and the estimated distance to the HVS, 
we trace back its trajectory in the Galactic potential.
Assuming it was ejected from the center, we find that a $\sim 2\masyr$
proper motion is necessary for the star to have come within a few parsecs 
of the SMBH.
We perform three-body scattering experiments to constrain the
progenitor encounter which accelerated the HVS.  
As proposed by Yu \& Tremaine (2003), we consider the tidal disruption
of binary systems by the SMBH and the encounter between a star
and a binary black hole, as well as an alternative scenario
involving intermediate mass black holes.
We find that the tidal disruption of a stellar binary ejects stars
with a larger velocity compared to the encounter between a single star and  
a binary black hole, but has a somewhat smaller ejection rate due to 
the greater availability of single stars.
\end{abstract}


\section{Introduction}
\label{sec:intro} 
The recent discovery by \citet{b05} of the hypervelocity star SDSS J090745.0+024507
(hereafter the HVS) has lent credence to the prediction of \citet{h88} that
dynamical encounters with the black hole in the Galactic centre
can eject stars with velocities up to several thousand kilometers per second. 
At a distance of 40-70$\kpc$, the HVS has a heliocentric radial velocity 
of $853 \pm 12\kms$ in a direction of $174^{\circ}$ from the Galactic centre 
\citep{b05}.
Once corrected for solar motion and galactic rotation, this translates into
a velocity of about $730\kms$ relative to the Local Standard of Rest. 
This velocity, which represents the lower limit of the total space velocity,
is significantly higher than that of any other runaway or high-velocity star 
in the Galaxy \citep{s91}.

We investigate the origin of the extreme velocity of the HVS
by means of kinematic analyses, binary evolution calculations and 
numerical scattering experiments.
There are three possible explanations for the velocity of the HVS:
(i) ejection upon supernova explosion of the companion in a binary system;
(ii) dynamical ejection after a close encounter with main sequence stars or
stellar mass compact objects;
(iii) dynamical ejection from the Galactic centre as a result of an encounter
with the supermassive black hole.

The paper is organised as follows: in \S2 we analyse the possibility
of an origin in the Galactic disk and an ejection by supernova explosion,
in \S3 we back-trace the orbit of the star in the Galactic potential
and constrain the value of its proper motion and in \S4 we describe
numerical scattering experiments of 3-body encounters involving
main sequence (MS) stars, the supermassive black hole (SMBH) 
and intermediate-mass black holes (IMBHs).

\section{The origin of the high velocity of the HVS}
\label{sec:origin}
To address the possibility that the HVS is a high-velocity runaway resulting
from the disintegration of a binary system, we turned to population
synthesis. Making use of the {\tt SeBa} stellar evolution package \citep{p01}, we
generated two sets of $10^6$ binaries each. In the first set, primary masses
ranged from $100\msun$ down to the hydrogen burning limit. 
In the second set, the minimum primary mass was increased to $8\msun$, in
order to get a larger sample of events for statistical purposes. 
The mass ratio in both cases was chosen from the distribution of
\citet{h92}, whilst the initial orbital separation was drawn from
that of \citet{d91} and truncated at $10^5$ R$_{\odot}$. 
The distribution of natal kick speeds for neutron stars was taken from
\citet{p90}, with a dispersion of $300\kms$. 
We assumed that the escape speed was the orbital speed of the secondary 
immediately before the binary disintegrated.
As well, we only considered secondary stars with a mass below $10\msun$, 
given the known constraints on the mass of the HVS. From this
synthesis experiment, we obtained maximum escape speeds of $\sim 70 \kms$, 
an order of magnitude below the space velocity of the HVS. 
Based upon this, we reject the high-velocity runaway hypothesis.
The possibility that the system is a binary and that its high speed is the
result of an asymmetric supernova kick can be discounted because, if we
assume that the visible companion is on the main sequence, a kick magnitude
in excess of $2000\kms$ would be required. While this can not be ruled out
physically, the probability of the system remaining bound in such an event
is negligible.

The only alternative explanation for the high velocity of the HVS
is an ejection during a dynamical encounter. 
Stellar encounters involving MS stars, neutron stars or stellar mass black 
holes eject stars with maximum velocities on the order of the orbital velocity 
in binary systems \citep{gpe04, sp93}. 
\citet{b05} assume therefore that the HVS was ejected from the 
Galactic centre by a strong encounter with the SMBH. 
This possibility was first proposed by \citet{h88}, who considered encounters 
of binaries with a SMBH, and then further developed by \citet{yt03}, 
who found that encounters between binaries and the SMBH or between
single stars and a binary black hole represent the most efficient channel 
to eject hypervelocity stars.

\section{The trajectory of the HVS in the Galaxy}
\label{sec:orbit}
\begin{figure}
\begin{center}
\includegraphics[width=7cm]{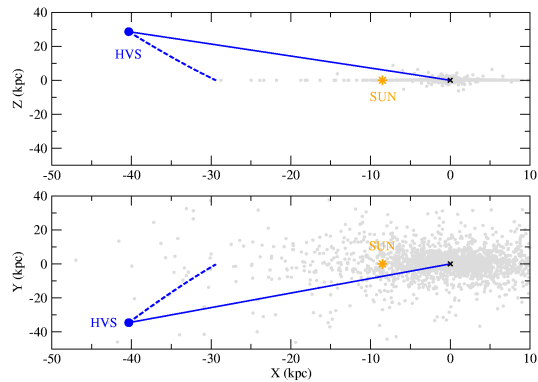}
\end{center}
\caption{Trajectory of the HVS integrated backward in time 
in the Galactic potential without any proper motion component (dashed line)
and with a proper motion of about 1.8$\masyr$ (solid line).
The integration is stopped upon passage through the disc. 
The full dot represents the present position of the star at a distance of
$55\kpc$ while the grey dots represent a schematic model of the Galaxy.}
\label{fig:orbit}
\end{figure}

In the previous section we argue that the HVS must have been
ejected from the Galactic centre by a dynamical encounter
with the SMBH. This implies that the HVS originated
in the Milky Way's central region.
Despite the incompleteness of available kinematic data, we trace back
the orbit of the HVS in the Galactic potential.
The distance of the object is still uncertain, as it depends on its
spectral type and evolutionary state.
\citet{b05} estimate a distance of $71\kpc$ for a B9.2 MS star and 
$39\kpc$ for a blue horizontal branch star, with an average value 
$d=55\kpc$. We consider all three values in our analysis.
We first assume that the star has no proper motion and trace its trajectory
backward in time until it crosses the disc, using Paczynski's model (1990) 
for the potential of the Galaxy. 
The orbit, shown in Fig. \ref{fig:orbit} (dashed line) for the case
$d=55\kpc$, doesn't pass through the Galactic centre region. 
We then randomly generate the two proper motion components
in the range 1-3$\masyr$ and derive the three-dimensional positions and
velocities in the galactocentric reference frame \citep{js87}.
For each distance value we generate about 5000 sets of initial conditions,
integrate the trajectories and determine the minimum distance $d_{\rm min}$ to the
Galactic centre. We find that there is at least one
combination of proper motion components for each distance such that 
$d_{\rm min}< 5\,\rm pc$.
In Tab. \ref{tab:pm} we report, for each distance $d$, the average values
$\mu_{\alpha}$ and $\mu_{\delta}$ of the proper motion components 
such that $d_{\rm min}<$10\,pc, the corresponding heliocentric velocity 
$V_{\rm SUN}$ and the velocity $V_{\rm  ej}$ at a distance $d_{\rm min}$ 
from the centre, corrected for galactic rotation.
\begin{table}
  \caption{Kinematic parameters of the HVS which minimize the distance
    from the Galactic centre. For each distance value, we report the average
    proper motion in right ascension $\mu_{\alpha}$ and declination $\mu_{\delta}$
    such that $d_{\rm min}<$10\,pc, the corresponding heliocentric velocity
    $V_{\rm SUN}$ and the velocity $V_{\rm ej}$ at a distance $d_{\rm min}$ 
    from the centre corrected for galactic rotation.}
  \label{tab:pm}
  \begin{center}
    \begin{tabular}{ccccc}
      \hline
      $d$ & $\mu_{\alpha}$ & $\mu_{\delta}$ & $V_{\rm SUN}$ & $V_{\rm ej}$ \\
      (kpc) & (mas/yr) & (mas/yr) & (km/s) & (km/s)\\
      \hline
      39 & -1.4916$\,\pm\,$0.0007 & 2.1679$\,\pm\,$0.0009 & 982$\,\pm\,$1 & 1246$\,\pm\,$144\\
      55 & -0.9769$\,\pm\,$0.0004 & 1.5379$\,\pm\,$0.0003 & 976$\,\pm\,$1 & 1266$\,\pm\,$66\\
      71 & -0.7214$\,\pm\,$0.0002 & 1.1917$\,\pm\,$0.0003 & 973$\,\pm\,$1 & 1249$\,\pm\,$142\\
      \hline
    \end{tabular}
  \end{center}
\end{table}
The ejection velocity  $V_{\rm  ej}$ is about $1250\kms$, independent of the 
assumed distance.
In the case $d=55\kpc$, the mean values $\mu_{\alpha}$ and $\mu_{\delta}$ given 
in Tab. \ref{tab:pm} result in the orbit shown in Fig. \ref{fig:orbit} (solid line).

We can use this analysis to calculate the minimum velocity
with which the HVS should have been ejected from the disk in the case
of the binary supernova scenario. Assuming a random proper motion in the range
1-3$\masyr$, the minimum velocity at disk crossing (corrected for galactic rotation)
is about $500\kms$, much larger than the average recoil velocity
from a supernova explosion (see \S~2). This result supports the scenario
of an origin in the Galactic centre.

\section{Three-body scatterings with the supermassive black hole}
\label{sec:scatter}
\begin{figure*}
\begin{center}
\includegraphics[width=5.5cm]{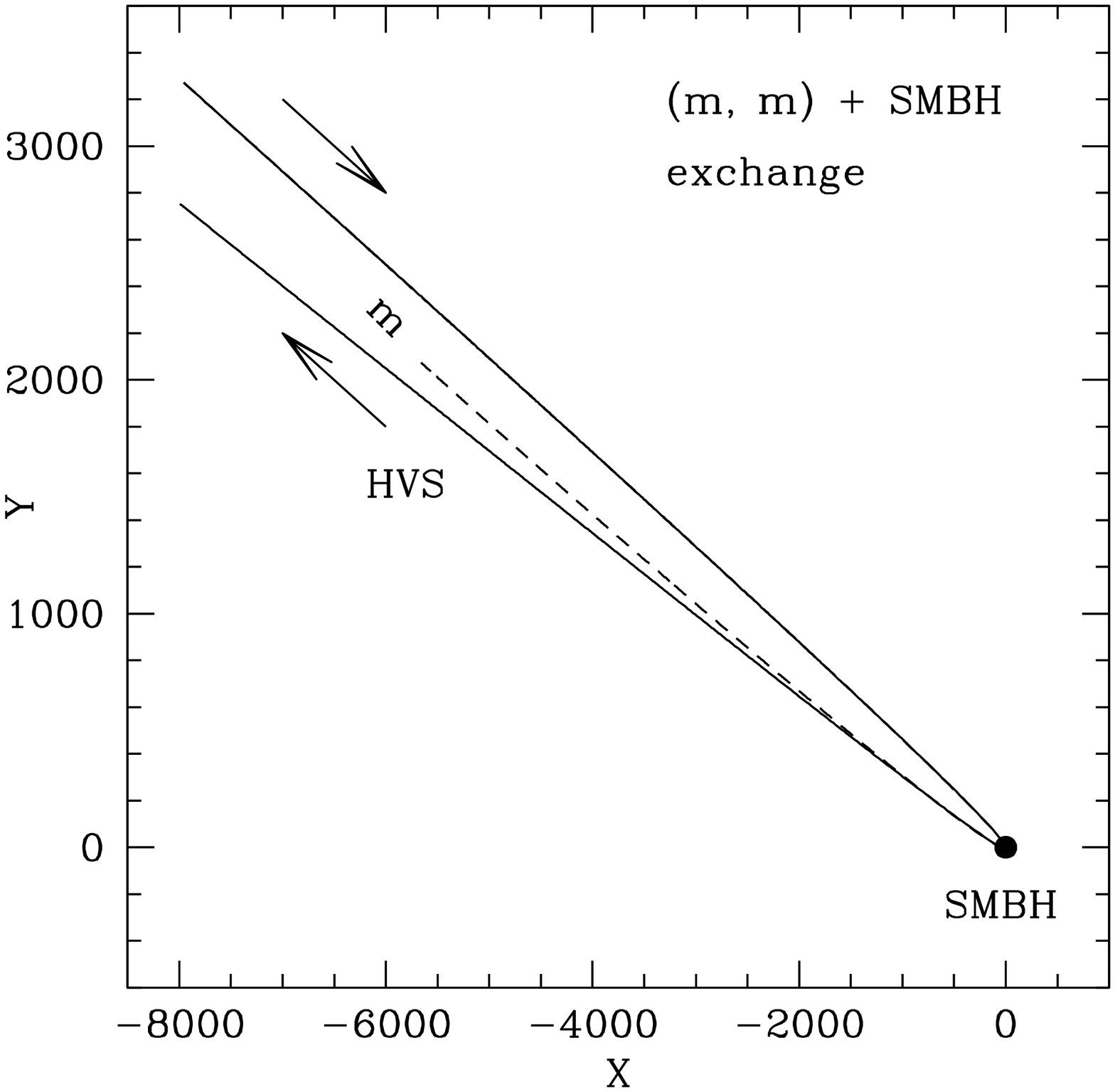}
\includegraphics[width=5.5cm]{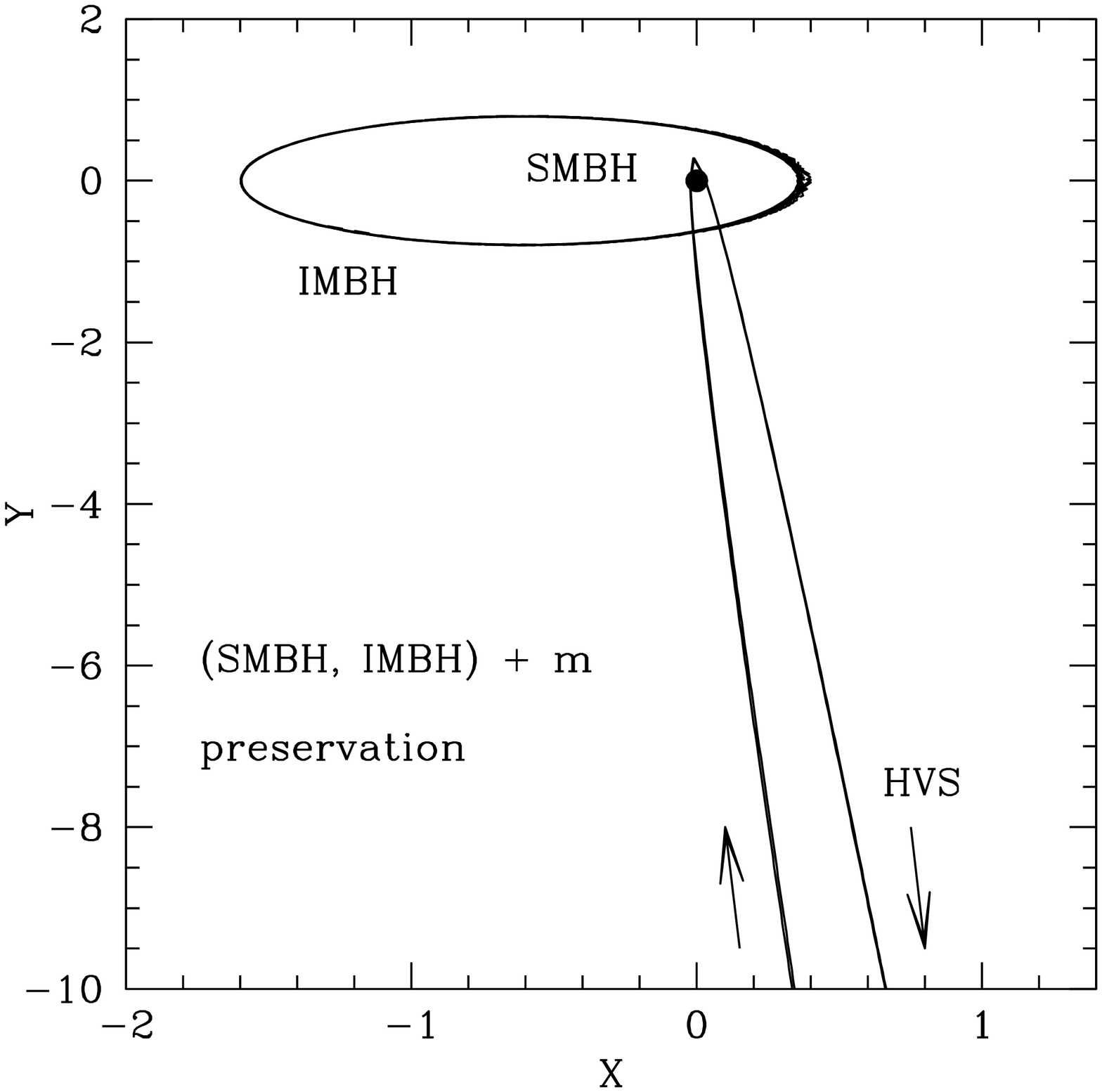}
\includegraphics[width=5.5cm]{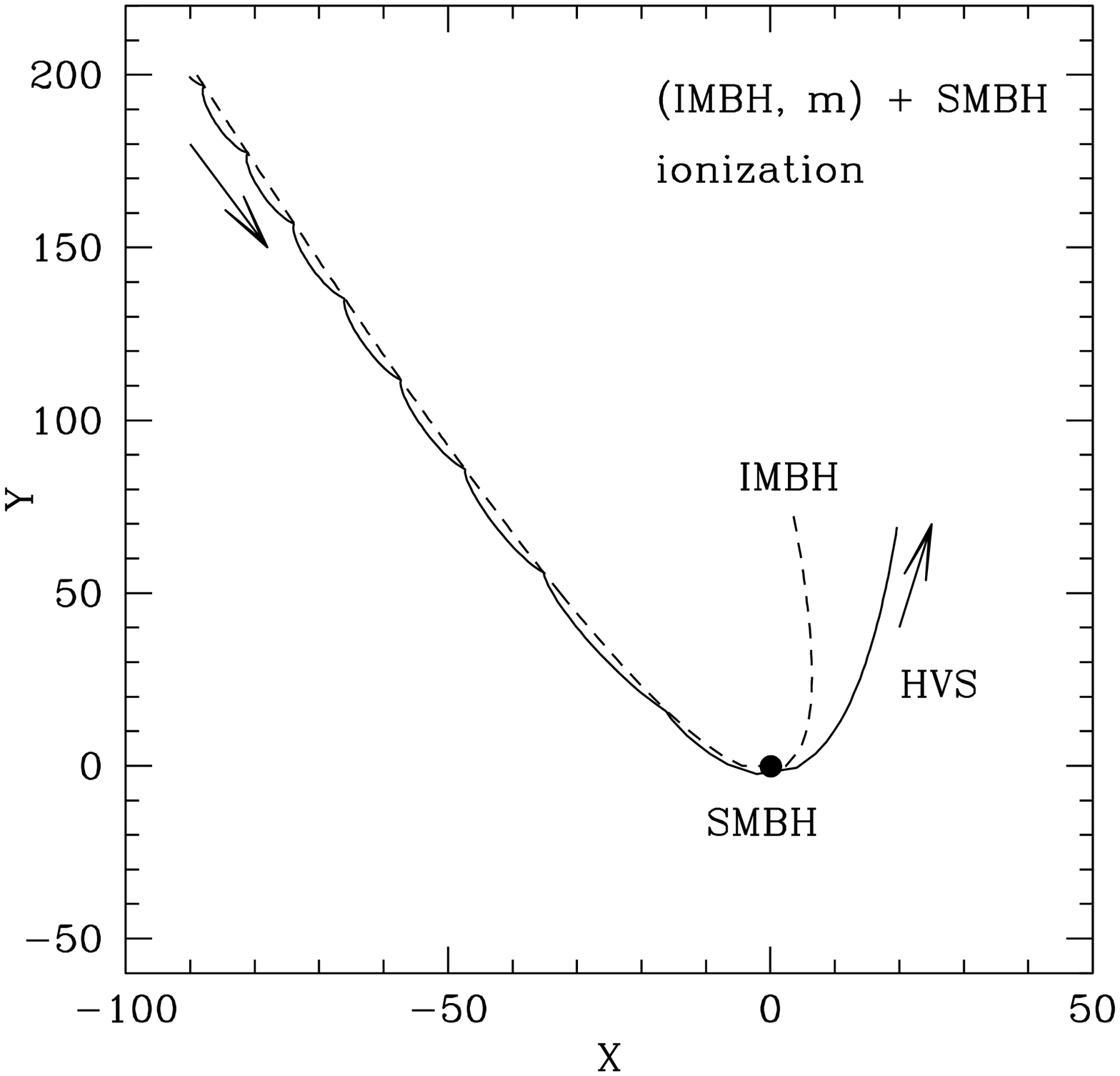}
\end{center}
\caption{Examples of three-body encounters involving the SMBH.
The positions are in units of the initial binary semi-major axis. 
(Left) An encounter between a binary of $3\msun$ MS stars and the
SMBH resulting in the the ejection of one star while the other star 
remains bound to the SMBH (exchange). 
(Middle) An encounter between a star and a black hole binary resulting in a 
preservation. 
(Right) An encounter between the SMBH and a binary containing a $3\msun$ 
MS star and an IMBH resulting in the breakup of the binary and the ejection 
of the MS star (ionization).}
\label{fig:enc}
\end{figure*}

We now explore the hypothesis of a dynamical ejection from the Galactic centre
by means of numerical simulations of three-body scatterings with the SMBH. 
In Fig. \ref{fig:enc} we show three examples of encounters
involving MS stars, the SMBH and an IMBH.
The experiments are carried out with the {\tt sigma3} package included in the
STARLAB\footnote{\tt http://www.manybody.org/manybody/starlab.html} software
environment \citep{mh96,p01}.  For each simulation we specify: the masses of
the three stars, the semi-major axis and eccentricity of the binary and the
relative velocity at infinity between the binary's centre of mass and the
single star. In order to classify possible collisions and mergers, physical
radii are also specified for the stars.  Additional parameters like the
orbital phase of the binary and its orientation relative to the incoming star
are randomly drawn from uniform distributions \citep{hb83}. The initial
eccentricity is drawn from a thermal distribution \citep{h75}.  The impact
parameter $b$ is randomized according to an equal probability distribution for
$b^2$ in the range $[0-b_{\rm max}]$.  The maximum value $b_{\rm max}$ is
determined automatically for each experiment (see \citet{gpe04} for a description).  
Energy conservation is usually better than one part in $10^6$ and, 
in case the error exceeds $10^{-5}$, the encounter is rejected.
The accuracy in the integrator is chosen in such a way that at most 5\% 
of the encounters are rejected. 

In all the experiments, we consider MS stars of mass
$m=3\msun$ (as indicated by \citet{b05} for a B9 star) and radius $R =
2.4\rsun$, an IMBH of mass $\mimbh=3000\msun$ and a SMBH of mass $\msmbh =
3.5\times10^6 \msun$ \citep{g03, s03}. The relative velocity at infinity
between the single star and the binary's centre of mass is set equal to the
dispersion velocity in the Galactic centre ($\sim 100\kms$).  The ejection
velocities of escapers are taken at the distance at which the integrator
stops \citep{mh96}. The effect of the black hole potential is negligible 
at this distance.

\subsection{Encounters between a binary of MS stars and the SMBH}
\label{sec:nb}
Close encounters between a binary and a very massive object can 
(i) break up the binaries and eject the two components with high speed 
({\it ionizations}), 
(ii) eject one star and leave the second star bound to the SMBH 
({\it exchange}) (see the left panel of Fig. \ref{fig:enc}).
We perform scattering experiments between the SMBH and binaries of MS stars. 
In all the runs, the binary stars have equal mass $m$
and the semi-major axis is varied in the range 0.05\,AU$ < a <$ 1\,AU. 

\begin{figure}
\begin{center}
\includegraphics[width=7cm]{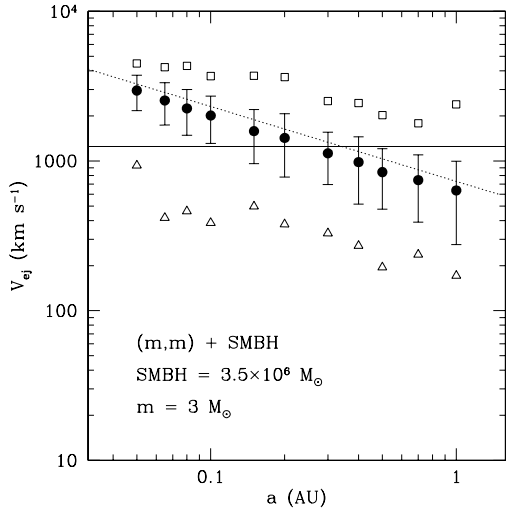}
\end{center}
\caption{Average recoil velocity of escapers as a function of the initial
  binary semi-major axis in the interaction of a stellar binary with the SMBH. 
  The error bars indicate the 2$\sigma$ deviation from the mean.
  The squares indicate the velocity $V_{\rm max}$ for which 1\% of the
  encounters have $V_{\rm ej} > V_{\rm max}$. The triangles indicate 
  the velocity $V_{\rm min}$ for which 1\% of the encounters have 
  $V_{\rm ej} < V_{\rm min}$. The horizontal line marks the $1250\kms$ 
  ejection velocity of the HVS while the dotted line gives the
  theoretical estimate by \citet{yt03}.}
\label{fig:nbvel}
\end{figure}

In Fig. \ref{fig:nbvel} we show the average ejection velocity $V_{\rm ej}$ 
of escapers as a function of the initial binary semi-major axis. 
Sufficiently high ejection velocities ($V_{\rm ej} \simgreat 1250\kms$) 
are obtained for $a\simless$ 0.3\,AU. The dotted line represents the
theoretical prediction by \citet{yt03} with an ejection speed parameter 
$v'_{\rm BH} = 130\kms$ (see Eq. 20).

\begin{figure}
\begin{center}
\includegraphics[width=7cm]{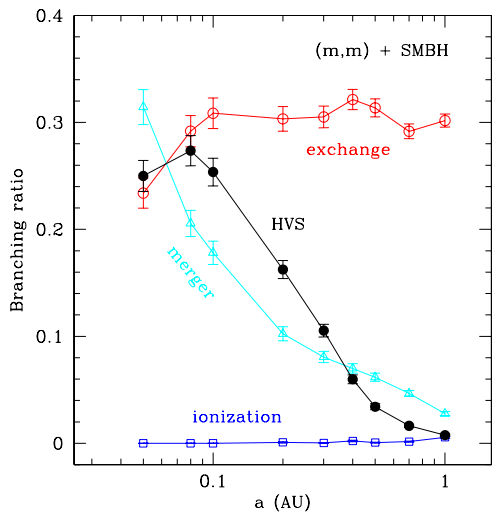}
\end{center}
\caption{Branching ratios as a function of the initial binary semi-major
  axis for encounters between a stellar binary and the SMBH.
  The empty symbols indicate the total fraction of ionizations,
  exchanges and mergers while the full dots indicate the fraction
  of encounters which produce escaping stars with a velocity larger 
  than $1250\kms$.}
\label{fig:nbbranch}
\end{figure}

In Fig. \ref{fig:nbbranch} we show the fraction of encounters resulting 
in ionization, exchange of the secondary star or merger for the range 
of $a$ under consideration.
The fraction of encounters resulting in an exchange of the secondary
star increases slowly from about 25\% for $a = 0.005$\,AU to about 30\% for
$a = 1$\, AU. Since the binary components have equal masses, 
the probability of ejection for the two stars is equal in exchange encounters.
Ionizations are rare and only occur for $a \simgreat$ 0.2\,AU, 
which marks the transition between hard and soft binaries \citep{h75}. 
As a result, high ejection velocities are mostly produced in exchange encounters.
The total fraction of scatterings whose outcome is a star escaping
with a velocity larger than $1250\kms$ decreases rapidly with increasing $a$,
and therefore binaries with semi-major axis 0.05\,AU $\simless a \simless$
0.3\,AU are the most suitable progenitors of hypervelocity stars.
If we consider that about 20\% of the simulated encounters
result in the ejection of a hypervelocity star over the suggested 
range of orbital separations, the ejection rate provided by \citet{yt03}
can be refined to $\sim 3\times10^{-6} \left(\eta/0.1\right) \rm yr^{-1}$,
where $\eta$ is the binary fraction.

The fraction of physical collisions and mergers decreases steadily with
increasing $a$ from $\sim$20\% to $\sim$2\%, and mainly involve the binary
components. Collisions with the SMBH occur in less than 1\% of the cases.
The merger products can remain bound to the SMBH or escape its gravitational potential.
We perform additional scattering experiments of encounters between binaries
and the SMBH to study the properties of the merger products.
We consider equal mass binaries with mass $m=1.5\msun$ (in such a way that the
mass of any possible merger is $\sim 3\msun$) and radius $R=1.4\rsun$.
Escapers are ejected with velocities larger than $1000\kms$ if the initial
semi-major axis is in the range 0.03\,AU $\simless a \simless$ 0.05\,AU. 
In this range, only about 6\% of the encounters result in a binary merger with
escape of the collision product. We therefore consider this scenario
inefficient for the production of hypervelocity stars.

\subsection{Encounters between a single star and a binary black hole}

We now consider the hypothesis that the SMBH is in a binary with an IMBH
and interacts with single stars (see the central panel of Fig. \ref{fig:enc}). 
The semi-major axis of the black hole binary is taken in the range 
2\,AU $< a < 1000$\,AU.
All the simulated encounters result in a preservation of the black hole
binary, during which the single star gains kinetic energy and escapes. 

\begin{figure}
\begin{center}
\includegraphics[width=7cm]{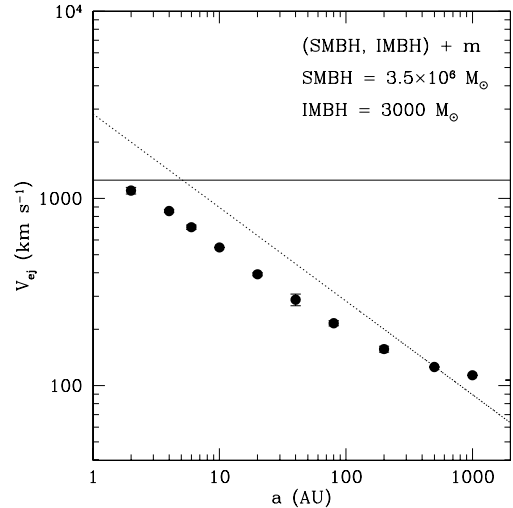}
\end{center}
\caption{Average recoil velocity of a single MS star escaping from 
  a black hole binary as a function of the binary's initial semi-major axis.
  The horizontal line marks the $1250\kms$ ejection velocity of the HVS while 
  the dotted line represents the analytical estimate by \citet{yt03}.}
\label{fig:bbhvel}
\end{figure}

In Fig. \ref{fig:bbhvel} we show the average ejection velocity 
of the escaping star as a function of $a$. 
The maximum velocity obtained in these encounters is about
$1000\kms$, barely sufficient to explain the velocity of the HVS. 
The comparison with the theoretical estimate (dotted line) by \citet{yt03} 
reveals a discrepancy of about a factor of 2 in the ejection velocities.
The numerical results also flatten at large initial semi-major axes and
deviate from the expected $V_{\rm ej} \propto a^{-1/2}$ scaling.
We have performed additional calculations with $1\msun$ incoming stars but
the average  $V_{\rm ej}$ appears to be insensitive to the mass of the single
star.

\subsection{Encounters between a SMBH and a main sequence star in a binary with an IMBH}
\label{sec:break}
\begin{figure}
\begin{center}
\includegraphics[width=7cm]{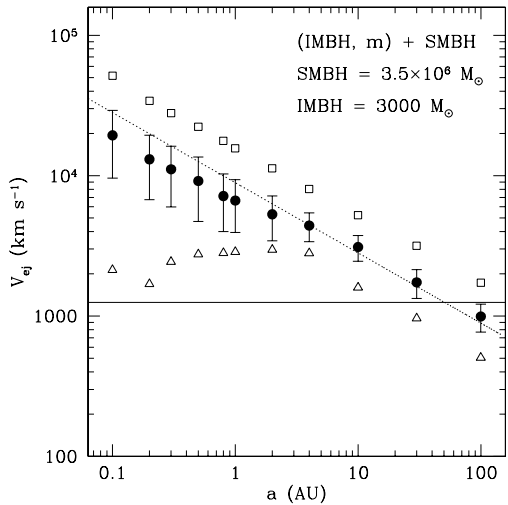}
\end{center}
\caption{Average recoil velocity of escapers as a function of the initial
  binary semi-major axis in the interaction between a normal star in a binary 
  with an IMBH and the SMBH. The error bars indicate a 2$\sigma$ deviation from the mean.
  The squares and triangles are defined as in Fig. \ref{fig:nbvel}. 
  The horizontal line marks the $1250\kms$ 
  ejection velocity of the HVS while the dotted line gives the 
  theoretical estimate by \citet{yt03}.}
\label{fig:bkvel}
\end{figure}

It has been proposed \citep{hm03} that the Galactic centre 
is populated by IMBHs, with masses in the range $100-10000\msun$.
These IMBHs may form in young dense star clusters as a
result of runaway collisions \citep{spz04} and sink toward the Galactic centre 
together with their parent cluster due to dynamical friction.
While the cluster dissolves in the Galactic potential, the IMBHs continue
to spiral in, possibly with a stellar companion, until they eventually
interact with the SMBH \citep{e01}. 
If this is the case, IMBHs must play a role in the dynamical encounters taking
place in the few inner parsecs of the Galaxy.
We consider encounters between the SMBH and a binary consisting of a MS star 
and an IMBH.
We select the binary semi-major axis in the range 0.1\,AU$< a < 100$\,AU, 
with the lower limit set by the IMBH's tidal radius 
$R_t = R \left(\mimbh/m\right)^{1/3} = 24\rsun$ for the adopted masses and radii. 
We focus on the encounters whose final outcome is the break-up of the binary
with subsequent ejection of star $m$ to infinity, while the IMBH
can either remain bound to the SMBH or escape (see the right panel of Fig. \ref{fig:enc}). 
Figure \ref{fig:bkvel} reports the velocity of the escaping star
after the encounter as a function of the initial $a$.
This type of encounter can easily eject stars with velocities of 
thousands of kilometers per second.
A theoretical estimate derived from \citet{yt03} (see Eq. 1--3) 
is shown with a dotted line. The numerical results agree well 
with the theoretical estimate.

\begin{figure}
\begin{center}
\includegraphics[width=7cm]{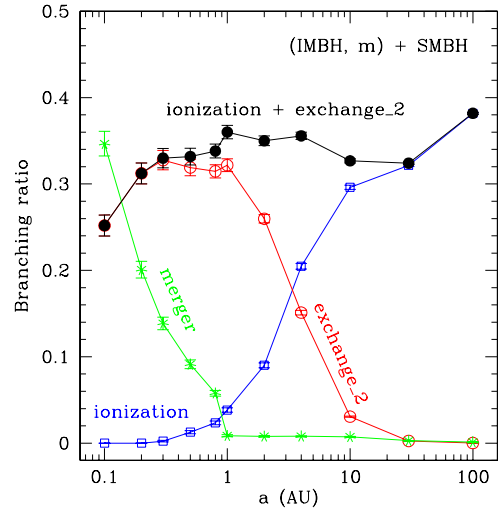}
\end{center}
\caption{Branching ratios as a function of the initial
  binary semi-major axis in the interaction between a normal star in a binary 
  with an IMBH and the SMBH.}
\label{fig:bkbranch}
\end{figure}

In Fig. \ref{fig:bkbranch} we show the fraction of encounters resulting 
in ionization, exchange of the secondary star or merger for the range 
of $a$ under consideration.
For $a \simless$ 0.3\,AU the binary is hard and the total cross-section is dominated 
by exchange encounters. In this case, the escaping star gains energy at the
expense of the IMBH, which becomes bound to the SMBH.
For larger semi-major axes, the binary is soft and tends to be ionized. 
The highest recoil velocities are generally obtained in exchange
encounters.

Although this scenario can eject stars with hypervelocities over a wide range
of orbital separations, only rarely eccentric orbits result in prompt
ionization.  Prompt ionization only occurs for low angular momentum
orbits, which comprise a fraction of about $10^{-5}$ if we assume an isotropic
velocity distribution.  Adopting a formation rate of $\sim 10^{-7} \rm
yr^{-1}$ for IMBHs \cite{spz05}, the ejection rate can be as small as 
$\eta 10^{-11} \rm yr^{-1}$, where $\eta$ represents the fraction of IMBHs 
with a stellar companion.  
Low eccentricity orbits result in a slow inspiral of the binary
due to dynamical friction, but, in this case, it is not clear whether 
the stellar density is sufficiently high to drag the binary to the tidal
radius of the SMBH. If an encounter with the SMBH does take place,
we expect the ejection velocity to be much lower than in the case of 
a prompt ionization.

\section{Summary and Conclusions}
\label{sec:disc}
We have investigated the origin of the extreme velocity of the HVS
by means of kinematic analyses, binary evolution calculations
and numerical simulations of three-body encounters.

By tracing the trajectory of the HVS in the Galactic potential 
(using the available measurements for the distance and the radial velocity),
we have shown that a proper motion of about $\sim 2\masyr$ is required 
for the star to have come within a few parsecs from the SMBH.
 
We confirm the prediction by \citet{h88} and \citet{yt03} that dynamical
encounters with the SMBH can eject hypervelocity stars to the Galactic halo;
the HVS is likely the first discovered object of this kind.
The most promising scenarios are dynamical encounters of MS stars with 
the SMBH, possibly in a binary with an IMBH.
In particular, the encounter between a SMBH and a stellar binary 
or between a single star and a binary black hole provide enough kinetic 
energy to eject hypervelocity stars over a restricted range of initial 
semi-major axes.
We have also investigated the more exotic encounter between a SMBH 
and an IMBH orbited by a stellar companion.
Although this type of encounter can be very energetic and doesn't require
any constraint on the binary's hardness, it has a very low probability. 
The possibility that the HVS is the product of a merger induced by the SMBH 
is intriguing but not very likely.

\section{Acknowledgments}
We thank Clovis Hopman and Melvin Davies for interesting discussions 
on hypervelocity stars and the anonymous referee for brief
and insightful comments on the manuscript.
This work was supported by the Netherlands Organization
for Scientific Research (NWO), the Royal Netherlands Academy
of Arts and Sciences (KNAW) and the Netherlands Research School 
for Astronomy (NOVA).

\end{document}